\documentclass[aps,pra,preprint,tightenlines,groupedaddress,showpacs]{revtex4}

\usepackage{graphics}
\begin{document}
\title{Collective modes of $p$--wave superfluid Fermi gases in BEC phase}
\author{F. Matera$^{1}$}
\email{matera@fi.infn.it}
\author{M. F. Wagner$^{2}$}
\email{mfwagner@mailaps.org}
\affiliation{
$^1${\small\it Dipartimento di Fisica e Astronomia, Universit\`a degli Studi
di Firenze, and\\
Istituto Nazionale di Fisica Nucleare, Sezione di Firenze\\
Via G. Sansone 1, I-50019, Sesto Fiorentino, Firenze, Italy}\\
$^2${\small\it  Frankfurt University of Applied Sciences\\
Nibelungenplatz 1, D--60318, Frankfurt am Main, Germany}}
\date{\today}

\begin{abstract}
The low--energy modes of a superfluid atomic Fermi gas at zero temperature are investigated. 
The Bose--Einstein--condensate (BEC)  side of the superfluid phase is studied in detail. 
The atoms are assumed to be in only one internal state, so that for a sufficiently diluted gas the pairing of fermions can be considered effective in the $l=1$ channel only. In agreement with previous works on $p$--wave superfluidity in Fermi systems, it is found that the $p_x+ip_y$ phase represents the lowest energy state in both the Bardeen--Cooper--Schrieffer (BCS) and BEC sides. 
Our calculations show that at low energy three branches of collective modes can emerge, with different species of dispersion relations: a phonon--like mode, a single--particle--like mode and a gapped mode. 
A comparison with the Bogoliubov excitations of the corresponding spinor Bose condensate is made. 
They reproduce the dispersion relations of the excitation modes of the $p$--wave superfluid Fermi gas to a high accuracy.  
\end{abstract}

\maketitle

\section{Introduction}
\label{intro}
Over the last several years a sustained interest has been devoted  to paired fermion phases with 
unconventional pairing symmetry. Among these the $p$--wave spin triplet condensate has attracted 
particular attention. The occurrence of $p$--wave superfluidity in Fermi systems was already 
studied in the sixties by Anderson and Morel \cite {An61} in relation to the low temperature phase of liquid $^3He$.  In the last decade there has been a renewed interest after the observation of 
$p$--wave Feshbach resonances in $^6Li$ and $^{40}K$ atoms \cite{Re03,Ti04,Zh04,Sc05}, which had  raised the prospect of realizing $p$--wave superfluidity in cold Fermi gases. In particular, the possibility 
of controlling the strength of the interatomic interactions via Feshbach resonances  
has stimulated theory works on the evolution of the superfluidity from the
Bardeen--Cooper--Schrieffer (BCS) regime to the Bose--Einstein
condensation (BEC)  of composite bosons  
\cite{Eng97,Gu05,Oh05,Ho05,Ch05,Is106,Is206,Bu06,Pa011,In013,Ca013,Li013}. We remark that
a $p$--wave superfluid phase may occur also in the Fermi component of a gaseous mixture of 
ultracold bosons and one--component (spin--polarized) fermions. In that case an attractive 
interaction between fermions can be induced by the exchange of virtual phonons \cite{Mat03,Mat11}.  
\par
In this work we explore properties of a three--dimensional spin--polarized Fermi gas  in  
the $p$--wave superfluid phase using a fermion--only model.  Since for the $p$--wave 
superfluid phase the order parameter is given by a three--dimensional object two 
distinct phases are available to the superfluid. A $p_x+ip_y$ phase, which 
is characterized by the $z$--component of the pair angular momentum $m$ equal to $\pm 1$ and a 
$p_x$ phase with $m=0$. At low temperatures the system can undergo a phase transition from a 
$p_x$ to a $p_x+ip_y$ superfluid state \cite{Gu05}. Moreover, detuning the Fesh-
bach resonance 
the system can exhibit a transition from a gapless quasiparticle spectrum for positive values of the 
chemical potential $\mu$  (BCS--side) to a gapped spectrum for $\mu<0$ (BEC--side)  \cite{Gu05,Is106}. We will address in particular the study of low--energy collective modes in the limit 
of vanishing temperature. The threshold for pair--breaking vanishes in the BCS--side. So the collective modes in general should be strongly damped in the BCS--side. For this reason, we will consider collective excitations of the Fermi superfluid in the BEC phase only. Collective modes in 
a $p$--wave superfluid Fermi gas have been already studied in Ref. \cite{Is206}. However, in that paper 
calculations were performed only for the $p_x$ superfluid phase. Moreover, the authors reported  dispersion relations only for the Nambu--Goldstone mode related to phase--oscillations of the pairing field. We aim to do a step further. We consider the system in the $p_x+ip_y$ phase, which represents the true ground state, and we extend calculations to collective modes given by 
amplitude fluctuations of the pairing field. Furthermore, we derive from the properties 
of the superfluid Fermi gas in the BEC side the structure of the corresponding Bose 
counterpart and the interaction between the composite bosons. Since the order parameter 
of the fermion superfluid phase is represented by a complex vector the boson gas can 
be considered as a Bose spinor condensate. We compare the dispersion relations for the 
collective modes of the Fermi system with the corresponding ones of the Bose condensate. From 
an experimental point of view the observation of the peculiarities of the excitation spectrum 
may help to assess the occurrence of a BEC regime in a $p$--wave superfluid Fermi gas \cite{Li14}.

\section{Formalism}
In the imaginary--time functional formalism an effective action for the pairing field can be 
obtained by using a Stratonovich--Hubbard transformation for the couples of fermion fields 
(see, e.g., Ref. \cite {Altland}). After integration of the fermionic part  the effective action 
in momentum representation is given by 
\begin{eqnarray}
S_{eff}(\Delta,\Delta^*)=&&-\frac{1}{2}\sum_{{\bf k},{\bf k}^{\prime},{\bf P}}\int^{\beta}_0d\tau
\Big[\Delta^*({\bf P},{\bf k},\tau)V^{-1}({\bf k},{\bf k}^{\prime})
\Delta({\bf P},{\bf k}^{\prime},\tau)\Big]
\\
\nonumber
&&-\frac{1}{2}\sum_{n}\frac{(-1)^{n+1}}{n}{\rm Tr}\,({\widehat{\cal G}}_0{\widehat \Delta})^n
\, ,
\label{aux}
\end{eqnarray}
where ${\widehat{\cal G}}_0$ is a $2\times 2$ diagonal matrix, whose 
elements are the imaginary--time propagators for independent fermions with 
chemical potential $\mu$, and the matrix 
${\widehat \Delta}$ 
\[{\widehat \Delta}({\bf P},{\bf k},\tau)
=\left( \begin{array} {cc}
               0 & \Delta({\bf P},{\bf k},\tau)\\
               \Delta^*({\bf P},-{\bf k},\tau) & 0 
                         \end{array}\right) \]
represents a complex bosonic field, which is periodic in the imaginary time interval ($0,\beta=1/T$) 
(units such that $\hbar=k_B=1$ are used).                           

In Eq.(\ref{aux}) the matrix  $V^{-1}({\bf k},{\bf k}^{\prime})$ 
is the inverse of the fermion--fermion interaction 
\begin{equation}
<{\bf P}/2-{\bf k},{\bf P}/2+{\bf k}|V|{\bf P}/2+{\bf k}^\prime,{\bf P}/2-{\bf k}^\prime)>
=\frac{1}{\cal V}V(|{\bf k}-{\bf k}^{\prime}|)\,
\label{pot}
\end{equation}
where ${\bf k}=({\bf k}_1-{\bf k}_2)/2$ and  ${\bf k}^{\prime}=({\bf k}_3-{\bf k}_4)/2$ are the relative momenta of a pair of fermions, and ${\bf P}={\bf k}_1+{\bf k}_2={\bf k}_3+{\bf k}_4$ is its 
center--of--mass momentum. In the above equation $V(|{\bf k}-{\bf k}^{\prime}|) $ represents the 
space--Fourier transform of the interaction and ${\cal V}$ is the normalization volume.  
\par
The pairing field at equilibrium, $\Delta_0({\bf P},{\bf k},\tau)$, is evaluated within the saddle--point approximation to the effective action $S_{eff}(\Delta,\Delta^*)$, while for the fluctuations of 
$\Delta({\bf P},{\bf k},\tau)$ about its equilibrium values  we adopt a gaussian approximation. The equation for  $\Delta_0({\bf P},{\bf k},\tau)$ reads  
 \begin{equation}
\Delta_0({\bf P},{\bf k},\tau)=
\label{pair1}
-\frac{1}{\cal V}\sum _{{\bf k}^{\prime}}V(|{\bf k}-{\bf k}^{\prime}|)
{\cal G}^{(12)}({\bf k}^{\prime}+{\bf P}/2,{\bf k}^{\prime}-{\bf P}/2,\tau-\tau^\prime=0)\,,
\end{equation}
where ${\cal G}^{(12)}({\bf k}^{\prime}+{\bf P}/2,{\bf k}^{\prime}-{\bf P}/2,\tau-\tau^\prime=0)$
is the equal--time anomalous propagator for fermions interacting with the pairing 
field ${\widehat \Delta}$ \cite{Walecka}. 
\par
The solutions of Eq. (\ref{pair1}) with ${\bf P}\not=0$, 
correspond to a breaking of the space--translational symmetry, 
i.e. the LOFF phase \cite{Loff}. Here, only solutions with vanishing 
center of mass momentum are considered. In this case the gap equation becomes 
\begin{equation}
\Delta_0({\bf k})=-\frac{1}{\cal V}\sum _{{\bf k}^{\prime}}V(|{\bf k}-{\bf k}^{\prime}|)
\frac{1}{\beta}\sum_n\frac{\Delta_0({\bf k}^{\prime})}
{\omega_n^2+\xi^{2}_{k^{\prime}}
+\Delta_0({\bf k}^{\prime})\Delta_0^*({\bf k}^{\prime})}\,.
\label{gap} 
\end{equation}
\par
Now, we expand the effective action $S_{eff}(\Delta,\Delta^*)$ up to the second order in the fluctuations of the pairing field 
\[\sigma({\bf P},{\bf k},\tau)=\Delta({\bf P},{\bf k},\tau)-\Delta_0({\bf k})\, \]
and for the effective action of the fluctuations we get
\begin{eqnarray}
&&\delta S_{eff}(\sigma,\sigma^*)=  -\frac{1}{4}\sum_{{\bf P},{\bf k},{\bf k}^{\prime}}\int_0^{\beta}d\tau
\Big[{\widehat\sigma}^\dag({\bf P},{\bf k},\tau){\widehat V}^{-1}({\bf k},{\bf k}^{\prime})
{\widehat\sigma}({\bf P},{\bf k}^{\prime},\tau)\Big]
\nonumber
\\
&&-\frac{1}{4}\sum_{{\bf P},{\bf k}}
\int_0^{\beta}d\tau d\tau^{\prime}\Big[{\widehat\sigma}^\dag({\bf P},{\bf k},\tau)
{\widehat A}(\tau-\tau^{\prime},{\bf P},{\bf k}) {\widehat\sigma}({\bf P},{\bf k},\tau^{\prime})\Big]\,, 
\label{gauss}
\end{eqnarray}
with the vector ${\widehat\sigma}({\bf P},{\bf k},\tau)$ and the matrix 
${\widehat V}^{-1}({\bf k},{\bf k}^{\prime})$ given by 
\[{\widehat \sigma}({\bf P},{\bf k},\tau)
=\left( \begin{array} {cc}
               \sigma({\bf P},{\bf k},\tau)\\
               \sigma^*({\bf P},{\bf k},\tau) 
                         \end{array}\right) \]                         
 and                        
\[{\widehat V}^{-1}({\bf k},{\bf k}^\prime)
=\left( \begin{array} {cc}
               V^{-1}({\bf k},{\bf k}^\prime) & 0 \\
           0 &  V^{-1}({\bf k},{\bf k}^{\prime}) 
                         \end{array}\right) \,.\]
In Eq.(\ref{gauss}) we have introduced the $2\times2$ matrix \\
${\widehat A}(\tau-\tau^{\prime},{\bf P},{\bf k})$ whose elements are 
\begin{eqnarray} 
A_{1,1}(\tau-\tau^{\prime},{\bf P},{\bf k})=
 -{\cal G}^{(22)}(\tau-\tau^{\prime},{\bf P}/2+{\bf k})
{\cal G}^{(11)}(\tau^{\prime}-\tau,{\bf P}/2-{\bf k})\,,
\nonumber
\\
A_{2,2}(\tau-\tau^{\prime},{\bf P},{\bf k})=
-{\cal G}^{(11)}(\tau-\tau^{\prime},{\bf P}/2+{\bf k})
{\cal G}^{(22)}(\tau^{\prime}-\tau,{\bf P}/2-{\bf k})\,,
\nonumber
\\
A_{1,2}(\tau-\tau^{\prime},{\bf P}/,{\bf k})=
{\cal G}^{(12)}(\tau-\tau^{\prime},{\bf P}/2+{\bf k})
{\cal G}^{(12)}(\tau-\tau^{\prime},{\bf P}/2-{\bf k})\,,
\nonumber
\\
A_{2,1}(\tau-\tau^{\prime},{\bf P},{\bf k})=
{\cal G}^{(21)}(\tau-\tau^{\prime},{\bf P}/2+{\bf k})
{\cal G}^{(21)}(\tau-\tau^{\prime},{\bf P}/2-{\bf k})\,,
\end{eqnarray}
where the  Green functions in the r.h.s. are calculated for the gas at equilibrium. Then, they are diagonal in the momenta. 
From general properties of the Green functions \cite{Walecka} the following symmetry 
relations ensue
\[
A_{2,2}(\tau-\tau^{\prime},{\bf P},{\bf k})=A_{1,1}(\tau^{\prime}-\tau,{\bf P},{\bf k})\]
and 
\[
A_{2,1}(\tau-\tau^{\prime},{\bf P},{\bf k})=A_{1,2}^*(\tau^{\prime}-\tau,-{\bf P},-{\bf k})\, .\]
\par
Equation  (\ref{gauss}) shows that the propagator for the bosonic field 
$[\sigma^*({\bf P},{\bf k},\tau),\sigma({\bf P},{\bf k},\tau)]^\dag$  
obeys the integral equation
\begin{eqnarray} 
{\widehat D}(\tau-\tau^\prime,{\bf P},{\bf k},{\bf k}^\prime)&&=\frac{2}{\cal V}
{\widehat V}(|{\bf k}-{\bf k}^\prime|)\delta(\tau-\tau^\prime)
\nonumber
\\
&&-\frac{1}{\cal V}\sum_{{\bf k}^"}\int_0^\beta d\tau_1\Big[{\widehat V}(|{\bf k}-{\bf k}^"|)
{\widehat A}(\tau-\tau_1,{\bf P},{\bf k}^"){\widehat D}(\tau_1-\tau^\prime,{\bf P},{\bf k}^",{\bf k}^\prime)\Big]\,.
\label{prop1}
\end{eqnarray}
\par
In order to calculate the dispersion relations of collective modes the above equations  should   analytically be continued  to real times. This can simply be accomplished by 
substituting in the expressions for the matrix elements $A_{i,j}$ 
the imaginary--time Green functions with their real--time counterparts . Then, we recast Eq.(\ref{prop1}) in the real--frequency representation  
\begin{equation}
{\widehat D}(\omega,{\bf P},{\bf k},{\bf k}^\prime)=\frac{2}{\cal V}{\widehat V}({\bf k},{\bf k}^\prime)
\label{prop}
-\frac{1}{\cal V}\sum_{{\bf k}^"}{\widehat V}({\bf k},{\bf k}^")
{\widehat A}(\omega,{\bf P},{\bf k}^"){\widehat D}(\omega,{\bf P},{\bf k}^",{\bf k}^\prime)\,.
\end{equation}
\par
Here, we consider a gas in the limit of  vanishing temperature $T=0$. In this
case the matrix elements  $A_{i,j}(\omega,{\bf P},{\bf k})$  are explicitly given by 
\begin{equation}
A_{1,1}(\omega,{\bf P},{\bf k})=\frac{u_+^2({\bf k})u_-^{2}({\bf k})}{\omega+E_+
({\bf k})+E_-({\bf k})-i\epsilon}
-\frac{v_+^2({\bf k})v_-^2({\bf k})}{\omega-E_+({\bf k})-E_-({\bf k})+i\epsilon}
\end{equation}
and 
\begin{equation}
A_{1,2}(\omega,{\bf P},{\bf k})=\frac{\Delta_+({\bf k})\Delta_-({\bf k})}{2E_+({\bf k})E_-({\bf k})}
\times\frac{E_+({\bf k})+E_-({\bf k})}{\omega^2-(E_+({\bf k})+E_-({\bf k})+i\epsilon)^2}\,,
\end{equation}
while the remaining matrix elements are determined by the symmetry relations 
\[A_{2,2}(\omega,{\bf P},{\bf k})=A_{1,1}(-\omega,{\bf P},{\bf k})\,,\]
\[A_{2,1}(\omega,{\bf P},{\bf k})=A_{1,2}^*(-\omega,-{\bf P},-{\bf k})\,.\]
In the above equations we have used the notations 
\[
u_{\pm}({\bf k}),v_{\pm}({\bf k}),\Delta_{\pm}({\bf k}),E_{\pm}({\bf k})=u({\bf P}/2\pm{\bf k}),
v({\bf P}/2\pm{\bf k}),
\Delta_0({\bf P}/2\pm{\bf k}),E({\bf P}/2\pm{\bf k})\, ,
\]
where  $E({\bf P}/2\pm{\bf k})=\sqrt{\xi^2({\bf P}/2\pm{\bf k})+|\Delta_0({\bf P}/2\pm{\bf k})|^2}$ is the quasiparticle energy, 
\[u^2({\bf P}/2\pm{\bf k})=\frac{1}{2}\Big(1+\frac{\xi({\bf P}/2\pm{\bf k})}{E({\bf P}/2\pm{\bf k})}\Big)\,,\]
and
\[v^2({\bf P}/2\pm{\bf k})=\frac{1}{2}\Big(1-\frac{\xi({\bf P}/2\pm{\bf k})}{E({\bf P}/2\pm{\bf k})}\Big)\,,\] 
with $\xi({\bf P}/2\pm{\bf k})=({\bf P}/2\pm{\bf k})^2/2m-\mu$. 
\par
The dispersion relations of elementary excitations are given by the real poles of the  propagator, Eq.(\ref{prop}), for given values of $\bf P$. In the vicinity of the poles the finite term $2{\widehat V}(|{\bf k},{\bf k}^\prime|)/{\cal V}$ in Eq.(\ref{prop}) can be neglected and the poles are determined by equating to zero   the determinant associated to the homogeneous set of equations 
\[
{\widehat D}(\omega,{\bf P},{\bf k},{\bf k}^\prime)=
-\frac{1}{\cal V}\sum_{{\bf k}^"}{\widehat V}({\bf k},{\bf k}^")
{\widehat A}(\omega,{\bf P},{\bf k}^"){\widehat D}(\omega,{\bf P},{\bf k}^",{\bf k}^\prime)\,.\]
The above equations admit solutions of the form 
\[{\widehat D}(\omega,{\bf P},{\bf k},{\bf k}^\prime)=
{\widehat \chi}_a(\omega,{\bf P},{\bf k})\times{\widehat \chi}_b(\omega,{\bf P},{\bf k}^\prime)\]
with the two--dimensional vector 
${\widehat \chi}(\omega,{\bf P},{\bf k})\equiv{\widehat \chi}_a(\omega,{\bf P},{\bf k})$ obeying the 
equations 
\begin{equation}
{\widehat \chi}(\omega,{\bf P},{\bf k})=
-\frac{1}{\cal V}\sum_{{\bf k}^\prime}{\widehat V}(|{\bf k}-{\bf k}^\prime)
{\widehat A}(\omega,{\bf P},{\bf k}^\prime){\widehat \chi}(\omega{\bf P},{\bf k}^\prime)\,.
\label{ampl}
\end{equation} 
\par
In the present work we are mainly interested in the global features of the collective modes. Then, a simplified interaction between fermions may be satisfactory for our purpose.  Firstly we assume that in the  expansion of the interaction in Legendre polynomials 
\[V(|{\bf k}-{\bf k}^\prime|)=\sum_l\sqrt{\frac{2l+1}{4\pi}}P_l(cos\theta_{\widehat {\bf k}{\bf k}^\prime})
V_l(k,k^\prime)\]
the term with $l=1$ gives the dominant contribution to the pairing of fermions. We observe that, 
because of the Pauli principle, only the components with odd values of the relative angular momentum $l$ can be effective. 
In addition, we adopt  for $V_1(k,k^\prime)$ the schematic form $V_1(k,k^\prime)=-\lambda kk^\prime$, with $\lambda>0$, for $k,k^\prime<k_c$, and  $V_1(k,k^\prime)=0$ otherwise. 
It is convenient to express the interaction 
strength $\lambda$ and the momentum cut--off $k_c$ in terms of  the scattering volume $v_1$ and the second coefficient, $r_1$,  in the effective--range expansion of the scattering amplitude \cite{Mot,Lan}. By exploiting the low energy expansion of the scattering matrix \cite{Ho05,In013,Pet} one can obtain the relations 
\begin{eqnarray} 
\frac{4\pi v_1}{m}&&= -\frac{\sqrt{{\displaystyle\frac{3}{4\pi}}}\lambda}{3-\sqrt{{\displaystyle\frac{3}{4\pi}}}{\displaystyle\frac{\lambda}{\cal V}\sum_{{\bf k}=0}^{k_c}
\frac{k^2}{2\epsilon_k}}}\,,
\nonumber
\\
&&\frac{1}{r_1}=-\frac{4\pi}{m^2}\frac{1}{\cal V}\sum_{{\bf k}=0}^{k_c}\frac{k^2}{2\epsilon _k^2}=-\frac{4}{\pi}k_c\,,
\label{para}
\end{eqnarray}
where $\epsilon_k=k^2/2m$. The diluteness condition $nr_1^3 <<1$ implies 
$(k_c/k_F)>>1$, where $n$ is the density of atoms and $k_F$ is the Fermi momentum. 
\par
The  equation (\ref{gap}) in the limit  $T=0$ becomes
\begin{equation}
\Delta_0({\bf k})=\lambda\sqrt{\frac{4\pi}{3}}\sum_mY_1^{m*}(\hat{\bf k})k\int \frac{d{\bf k}^\prime}{(2\pi)^3}k^\prime 
Y_1^m(\hat{\bf k}^\prime)\frac{\Delta_0({\bf k}^\prime)}{2E({\bf k}^\prime)}
\, ,
\label{pair2}
\end{equation}
and, in order to determine explicity the field $\Delta_0({\bf k})$ together with the chemical 
potential $\mu$  the equation fixing the fermion density has to be added
\begin{equation}
n_F=\int \frac{d{\bf k}}{(2\pi)^3}\frac{1}{2}
\left(1-\frac{\xi(k)}{E({\bf k)}}\right)\,,
\label{chim}
\end{equation}
\par
The energy per fermion in the superfluid phase is given by the expression 
\begin{equation}
E_F=\frac{1}{n_F}\int \frac{d{\bf k}}{(2\pi)^3}\frac{1}{2}
\left(\xi(k)-E({\bf k})+\frac{|\Delta_0({\bf k})|^2}
 {2 E({\bf k})}\right)+\mu\, ,
\label{ef}
\end{equation}
which coincides with the usual expression of the BCS theory
\cite{Walecka}, apart from a factor $1/2$ due to
the absence of degeneracy for the Fermi gas in the present case. 
\par
Equation (\ref{pair2}) suggests solutions of the form 
$\Delta_0({\bf k})={\bf \Delta}_0\cdot{\bf k}\theta(k_c-k)$, 
where the vector ${\bf \Delta}_0$ can be of two different species \cite{An61,Gu05,Ho05}: 
a real vector ($p_x$--phase) or a complex vector ($p_x+ip_y$--phase). In the latter case 
${\bf \Delta}_0 $ can be decomposed according to ${\bf \Delta}_0={\bf \Delta}_1+i{\bf \Delta}_2$ 
where ${\bf \Delta}_1$ and ${\bf \Delta}_2$ are two real vectors of equal magnitude and each other orthogonal. Moreover, vectors obtained by global gauge transformations and/or by rotations, in the orbital space, of a particular choice of ${\bf \Delta}_0$ represent equivalent solutions for the equilibrium pairing field. Explicit calculations show that the $p_x+ip_y$ phase is the lowest--energy state   \cite{An61,Gu05,Ho05}. 
\par
With the adopted interaction between fermions,  Eq.(\ref{ampl}) becomes 
\[
{\widehat \chi}(\omega,{\bf P},{\bf k})=\lambda\sqrt{\frac{4\pi}{3}}\sum_mY_1^{m*}(\hat{\bf k})k
\int \frac{d{\bf k}^\prime}{(2\pi)^3}k^\prime Y_1^m(\hat{\bf k}^\prime) 
{\widehat A}(\omega,{\bf P},{\bf k}^\prime){\widehat \chi}(\omega,{\bf P},{\bf k}^\prime)\,.
\]
The amplitudes ${\widehat \chi}(\omega,{\bf P},{\bf k})$ can be conveniently expanded as 
\[{\widehat \chi}(\omega,{\bf P},{\bf k})=\sum_m{\widehat \chi}^{(m)}(\omega,{\bf P})kY_1^{m*}(\hat{\bf k})\,,\]
and we obtain for the $m$--components the set of coupled equations 
\begin{equation}
{\widehat \chi}^{(m)}(\omega,{\bf P})=\sum_{m^\prime}\bigg[\lambda\sqrt{\frac{4\pi}{3}}
\int \frac{d{\bf k}^\prime}{(2\pi)^3}k^{\prime 2} Y_1^{m}(\hat{\bf k}^\prime) 
{\widehat A}(\omega,{\bf P},{\bf k}^\prime)
Y_1^{m^{\prime}*}(\hat{\bf k}^\prime)\bigg]{\widehat \chi}^{(m^\prime)}(\omega,{\bf P})\,.
\label{amplm}
\end{equation} 
\section{Results}
\label{sec:3}
In this paper we are concerned with the fluctuations of the pairing field in the stable phase 
$p_x+ip_y$. The ground--state energy  and the chemical potential are determined by the magnitude 
$\Delta_0=\sqrt {{\bf \Delta}_0\cdot{\bf \Delta}_0^*}$ alone. In order to calculate $\Delta_0$ we choose a 
particular direction for the vectors ${\bf\Delta}_1$ and ${\bf\Delta}_2$, say  
${\bf\Delta}_1=(\Delta_0/\sqrt{2},0,0)$ and  ${\bf\Delta}_2=(0,\Delta_0/\sqrt{2},0)$.  
Moreover, we observe that once the value of the cut--off $k_c$ is fixed, the scaled quantity 
${\tilde\Delta}_0=\Delta_0k_F/\epsilon_F$, where 
$\epsilon_F=k_F^2/2m$ is the Fermi energy, depends only on the dimensionless effective coupling constant 
\[{\tilde \lambda}=2m\lambda k_F^3\,,\]
which is a function of the product $v_1k_F^3$ and of the scaled cut--off $k_c/k_f$. 
In terms of the parameter $\tilde\lambda$ the equation for ${\tilde\Delta}_0$ reads 
\begin{equation}
1=\sqrt{\frac{3}{4\pi}}{\tilde\lambda}\int \frac{d{\tilde {\bf k}}}{(2\pi)^3}
\frac{1}{2}{\tilde k}^2sin^2(\theta_{\tilde{\bf k}})
\times\frac{\theta({\tilde k}_c-{\tilde k})}{2\sqrt{{\tilde\xi}^2({\tilde k})
+\frac{1}{2}{\tilde\Delta}_0^2{\tilde k}^2 sin^2(\theta_{\tilde{\bf k}})}}
\, ,
\label{pair}
\end{equation}
with ${\tilde\xi}({\tilde k})=\xi({\tilde k})/\epsilon_F$ and momenta expressed in units of the Fermi 
momentum: ${\tilde k},{\tilde k}_c=k/k_F,k_c/k_F$. \par
The dispersion relations for collective modes, $\omega({\bf P})$, are given by the frequencies, 
for which the determinant of the matrix
\[
{\widehat 1}-{\widehat A}_T(\omega,{\bf P})\equiv \delta_{m,m^\prime}\delta_{i,j}-
 A^{(m,m^\prime)}_{T(i,j)}(\omega,{\bf P})\,,\]
with  
\begin{equation}
A^{(m,m^\prime)}_{T(i,j)}(\omega,{\bf P})=\lambda\sqrt{\frac{4\pi}{3}}
\int \frac{d{\bf k}}{(2\pi)^3}\bigg[k^2Y_1^{m}(\hat{\bf k}) 
A_{i,j}(\omega,{\bf P},{\bf k})Y_1^{m^\prime*}(\hat{\bf k})\bigg] \,,
\label{matrix}
\end{equation}
vanishes for given values of the momentum ${\bf P}$. Similarly to the energy gap the scaled 
frequency $\omega({\bf P})/\epsilon_F$ depends only on the dimensionless parameter 
${\tilde\lambda}$ and the scaled cut--off ${\tilde k}_c$. 
\par
With the particular choice for the plane where the vector ${\bf\Delta}_0$ lies and for the 
phase of its components, the spatial $SO(3)$ symmetry and the gauge $U(1)$ symmetry are broken. We notice that in the present case the spins of the fermions do not play any role since they are frozen along 
a fixed direction and there is no coupling between spin and orbital degrees of freedom.  
As a consequence, we expect the occurrence of  Nambu--Goldstone modes related both to the 
phase fluctuations of the $m=1$ component of the pairing field and  to the fluctuations of 
the amplitudes of the pairing field.  Actually, by exploiting the gap equation  
one can see that Eqs. (\ref {amplm}) show solutions with ${\omega}=0$ for vanishing momentum 
 $P$. However, this does not give rise in general to undamped phonon--like excitations. This occurs 
because for particular directions of  the momentum the energy gap vanishes, so that in the 
BCS side the phonon--like modes can merge in the continuum of ungapped two--quasiparticle excitations.  
\par
\begin{figure}
\includegraphics{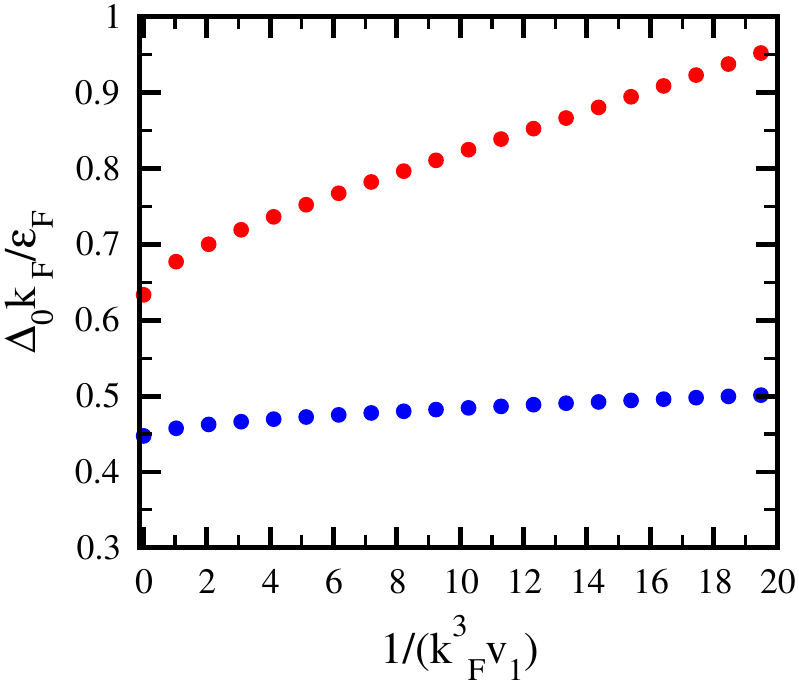}
\caption{The scaled value of the $p$--wave energy gap as a function of $1/(v_1k_F^3)$ for two values of the momentum cut--off: $k_c/k_F=10$ (red circles) and $k_c/k_F=20$ (blue circles).}
\label{fig1}
\end{figure}
\par
The threshold for the quasiparticle continuum is given by the minimum value of the sum 
$E({\bf P}/2+{\bf k}))+E({\bf P}/2-{\bf k})$. The most favorable case is when ${\bf P}$ is "parallel" to the 
complex vector ${\bf\Delta}_0$. In this case the threshold is given by $\Delta_0P/\sqrt{2}$, and only phonon--like modes with a phase velocity such that $v_S/v_F\leq {\tilde\Delta}_0/(2)^{3/2}$, 
where $v_F$ is the Fermi velocity,  can propagate without damping. Explicit calculations show that  this requirement can be satisfied only for large energy gaps. The corresponding values of the interaction strength are  in the region close to the BCS--BEC transition. We should point out that our calculations are essentially performed within the framework of a mean--field theory. The extension of the present approach  to a critical region, where fluctuations of the pairing field could be important, can be rather questionable. Therefore, we do not discuss collective modes in the BCS side and focus our attention on the collective excitations of the Fermi gas  inside the BEC phase. 
\par
The scaled energy gap ${\tilde\Delta}_0$  and the chemical potential together with the 
energy per particle are reported in Figs. (1) and (2) respectively as functions of the parameter  
$1/(v_1k_F^3)$ for two different values of the cut--off,  ${\tilde k}_c=10,20$. The curves for 
${\tilde k}_c=10$ show a steeper behavior. This is simply due to the fact that the effective coupling constant ${\tilde\lambda}$ is larger for smaller ${\tilde k}_c$ at a given value of $1/(v_1k_F^3)$.  
Figure (2) shows a remarkable closeness between the chemical potential and the energy per 
particle, the two sets of points are practically indistinguishable. This indicates that the pairs of fermions 
behave as a weakly interacting Bose condensate \cite{Leg}. 
\par
\begin{figure}
\includegraphics{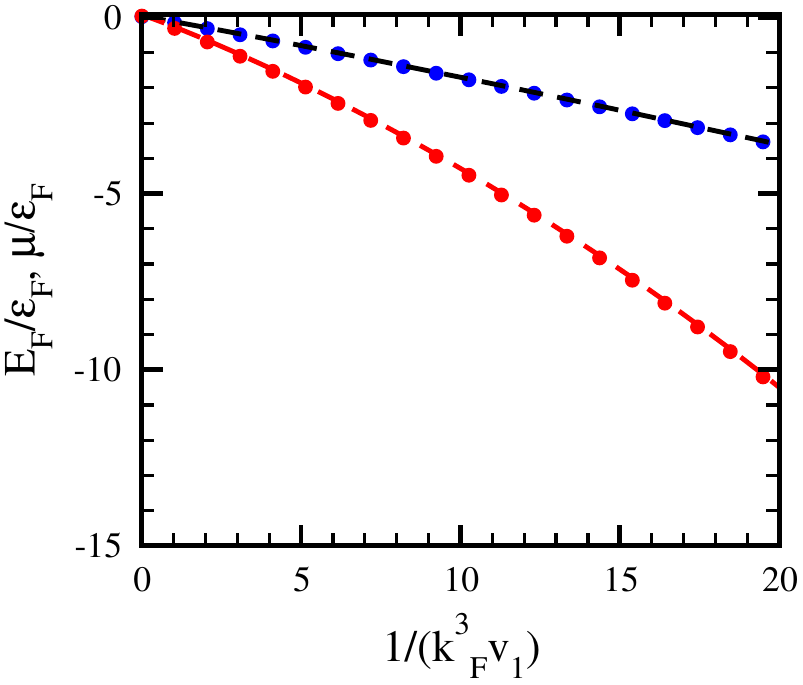}
\caption{ Energy per particle (circles) and chemical potential (dashed lines), in units of the Fermi energy, as a function of $1/(v_1k_F^3)$ for two values of the momentum cut--off: $k_c/k_F=10$ (red color) and $k_c/k_F=20$ (blue color).}
\label{fig2}
\end{figure}
Now we turn to  the dispersion relations of collective modes. In order to bring out any anisotropy in the 
propagation of the pairing--field fluctuations we consider two directions of the 
wave vector, orthogonal each other: ${\bf P}$ directed along the ${\bf z}$ axis, i.e. perpendicular to the 
complex vector ${\bf\Delta}_0$, and ${\bf P}$ lying in the $({\bf x},{\bf y})$ plane, parallel to the 
${\bf x}$ axis for definiteness. 
\par
In the first case, the quasiparticle energies $E({\bf P}/2\pm{\bf k})$ do not depend on the azimuthal angle, $\phi$, of ${\bf k}$. Then, the diagonal matrix elements $A_{i,i}(\omega, {\bf P},{\bf k})$ are 
independent of $\phi$ as well. Whereas the off--diagonal elements $A_{1,2}(\omega, {\bf P},{\bf k})$ 
and $A_{2,1}(\omega, {\bf P},{\bf k})$  are multiplied by the factors $e^{2i\phi}$ and $e^{-2i\phi}$ 
respectively. As a consequence, the amplitude with $m=0$ is decoupled from 
the remaining amplitudes. In addition, $\chi^{(1)}_1(\omega,{\bf P})$ is coupled only to 
$\chi^{(-1)}_2(\omega,{\bf P})$ and  the amplitudes $\chi^{(-1)}_1(\omega,{\bf P})$ 
and $\chi^{(1)}_2(\omega,{\bf P})$ separately obey single equations. 
In the other case, where ${\bf P}$ is parallel to the 
${\bf x}$ axis, the situation is quite different. The amplitude with $m=0$ is still independent of the 
remaining ones. Whereas, for the amplitudes with $m=\pm1$ we have to solve four coupled equations. 
However, explicit calculations show that  the matrix elements, which couple 
$\chi^{(m)}_i(\omega,{\bf P})$ to $\chi^{(-m)}_i(\omega,{\bf P})$ and $\chi^{(m)}_1(\omega,{\bf P})$  
to $\chi^{(m)}_2(\omega,{\bf P})$,  are about three orders of magnitude smaller than the other ones. 
Practically, the structure of the set of equations is the same in both the cases, but the values of the coefficients are different in general. 
\par
For the effective coupling constant we fix the value corresponding to  $v_1k_F^3=0.1$ 
and consider two different values of the scaled cut--off ${\tilde k}_c=20$ and 
${\tilde k}_c=10$. The values, in units of $\epsilon_F$, of the energy gap, 
chemical potential and energy per particle are 
\[{\tilde\Delta}_0=0.484, \qquad \mu=-1.70,\qquad \frac{E}{N}=-1.72\]
for ${\tilde k}_c=20$ and 
\[{\tilde\Delta}_0=0.821, \qquad \mu=-4.28,\qquad \frac{E}{N}=-4.34\]
for ${\tilde k}_c=10$. 
\par
\begin{figure}
\includegraphics{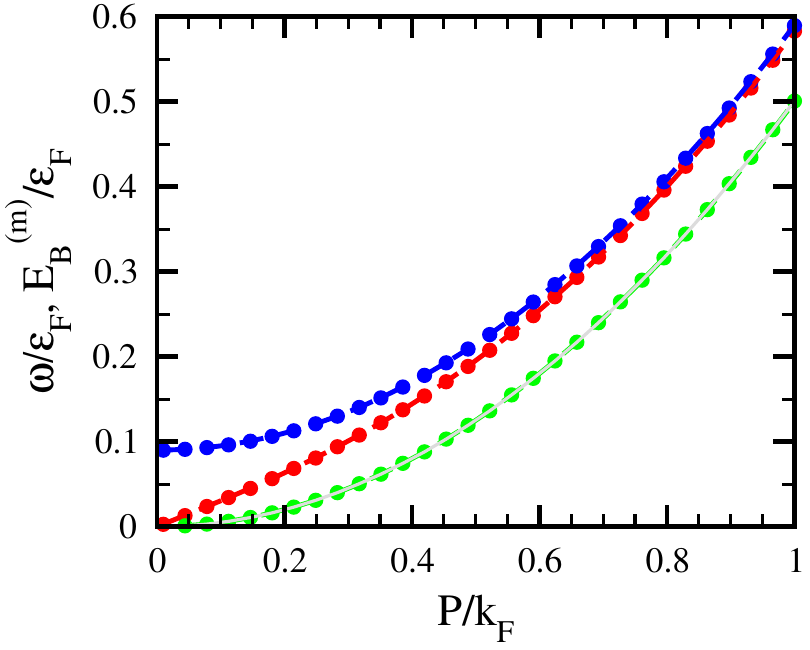}
\caption{Dispersion relations of collective modes of the superfluid Fermi gas
  (circles) and excitations energies of the corresponding Bose condensate
  (dashed lines) as a function of the scaled wave vector for
  $1/(k_F^3v_1)=10$. The scaled momentum cut--off is $k_c/k_F=20$.  
Red, blue and green colors correspond to $m=1,-1,0$ oscillations respectively (see text). }
\label{fig3}
\end{figure}
\begin{figure}
\includegraphics{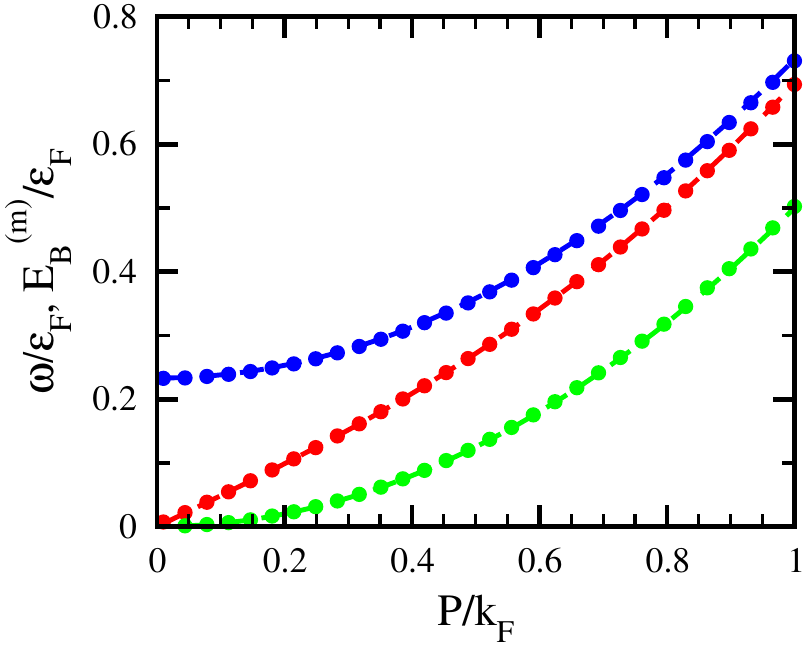}
\caption{Same as Fig. 3, but for $k_c/k_F=10$.  }
\label{fig4}
\end{figure}
\par
In Figs. (3) and (4) the dispersion relations $\omega({\bf P})$ are reported for the two cases considered and for modes propagating along a direction perpendicular to the equilibrium pairing field 
(${\bf z}$ axis). For modes propagating along the ${\bf x}$ axis the dispersion relations are very similar, 
 the sets of points are indistinguishable from those of Figs. (3) and (4).  Practically the propagation of the collective modes can be considered isotropic, the difference between the phase velocities for two orthogonal directions amounts to less than $1\%$ on average. 
 \par
 Figs. (3) and (4) show some remarkable features of the spectra of collective excitations. In 
 addition to the usual ungapped phonon--like branch related to phase fluctuations of the equilibrium  pairing field ($m=1$),  there are a gapped spectrum for the $m=-1$ amplitude and a parabolic  dispersion relation for the $m=0$ amplitude. 
From these properties of the dispersion relations we can gain some useful insight on the structure of the BEC and the interaction between composite bosons. The wave--function of the relative motion of the couple of fermions in the BEC side is an eigen--function of the relative angular momentum with $l=1$ and $m=0,\pm 1$ \cite{Mat11}. If the details of the internal structure are neglected the composite bosons can  be considered point--like particles with spin $S=1$. We have studied their equilibrium 
state and their dynamics by means of the time--independent and 
time--dependent  Gross--Pitaevskii equations (see, e.g., Ref. \cite{Pet}). 
Since the spectrum of elementary excitations is isotropic in orbital space, it is not necessary to 
introduce any coupling between spin and orbital degrees of freedom. 
The simple  interaction 
\begin{equation}
<{\bf q}_1,m_1;{\bf q}_2,m_2|\hat {V}|{\bf q}_3,m_3;{\bf q}_4,m_4>=
\delta_{{\bf q}_1+{\bf q}_2,{\bf q}_3
+{\bf q}_4}(\gamma_0+ \gamma_1<m_1,m_2|{\bf S}_1\cdot{\bf S}_2|m_3,m_4>)
\nonumber
\end{equation}
has been used. For an extensive review of spinor Bose--condensates see, e.g., Ref. \cite{Ka12}. 
\par
Since the Fermi system under consideration is in a $p_x+ip_y$ superfluid phase, we should expect that the corresponding Bose field at equilibrium, $\Phi_m^{(c)}$,  contains only the $m=1$ component 
(~ferromagnetic phase~), which can be  assumed real and positive.  In this phase,  from the time--independent  Gross--Pitaevskii equation one obtains for the difference between the chemical potential and energy per particle of the composite bosons the relation
\begin{equation} 
\mu_B-E/N_B=\frac{1}{2}(\gamma_0+\gamma_1)\Phi_1^{(c)2}\, ,
\label{gamma}
\end{equation} 
with $\mu_B=2\mu$,  $E/N_B=2E/N$ and $\Phi_1^{(c)2}=n_B=n/2 $. 
The requirement that $\Phi_1^{(c)}$ represents a state of stable equilibrium, implies that the 
inequalities  $\gamma_1<0$  and 
$\gamma_0+\gamma_1>0 $ should be fulfilled \cite{Ka12}. 
\par
As we expect a weak interaction between composite bosons, the Bogoliubov approximation to the 
solutions of the time--dependent Gross--Pitaevskij may be satisfactory. Within this approximation one obtains the following dispersion relations for the eigen--modes of the spinor Bose condensate 
\cite{Ka12}: a Nambu--Goldstone mode 
\[E_B^{(1)}(P)=\sqrt{\epsilon_B(P)[\epsilon_B(P)+2(\gamma_0+\gamma_1)n_B]}\,,\]
a single--particle--like mode 
\[E_B^{(0)}(P)=\epsilon_B(P)\]
and a gapped mode 
\[E_B^{(-1)}(P)=\epsilon_B(P)-2\gamma_1n_B\,,\] 
where $\epsilon_B(P)=P^2/(2M_B)$ with $M_B=2m$ is the boson kinetic energy.  
\par
The scaled energies $E_B^{(m)}(P)/\epsilon_F$ depend on the coupling constants only through the 
dimensionless quantities $\Gamma_i=\gamma_in_B/\epsilon_F$, with $i=0,1$. The values of $\Gamma_i$ can be determined  from Eq. (\ref{gamma}) and from the gap observed in the dispersion relation for the $m=-1$ fluctuations, Figs. (3) and (4). They are 
 \[\Gamma_0= 0.135,\qquad \Gamma_1=-0.045\]
 for ${\tilde k}_c=20$ and 
 \[\Gamma_0= 0.349,\qquad \Gamma_1=-0.116\] 
 for ${\tilde k}_c=10$.  
 \par
In Figs. (3) and (4) the dispersion relations  for the eigen--modes of the Bose condensate have been 
added. We observe that they reproduce the corresponding excitation 
spectra of the superfluid Fermi gas with noticeable accuracy. The phase--velocity of the phonon--like mode, in units of the Fermi velocity, is $0.150$ for ${\tilde k}_c=20$ and $0.241$ for ${\tilde k}_c=10$.  
 \par
 A remark is required about the symmetry properties of both the Fermi
 superfluid and the Bose spinor--condensate counterpart. The superfluid
 $p_x+ip_y$ phase and the corresponding ferromagnetic phase  
 for the BEC break both the rotational symmetry and the $U(1)$ symmetry under
 global gauge transformations. However,  $U(1)$ transformations can make up
 for spatial rotations about the original  
 ${\bf z}$ axis, i.e. the ground state displays a "spin--gauge" symmetry \cite{Oh98,Ho98,Uc10}. So that a $U(1)$ symmetry is still maintained. This reduces to $3$ the number of spontaneously broken group--generators. As a consequence $3$ Nambu--Goldstone modes should occur. Our calculations show a 
 phonon--like mode and a single--particle--like mode, in addition to a gapped mode. According to Ref. \cite{Ni76} the first mode should be counted once, whereas the second should be counted twice. Then, 
 the Goldstone theorem is still fulfilled.  

\section{Summary and conclusions}
\label{summary}
For a sufficiently diluted atomic Fermi gas with only one component,  
the interaction between atoms can be considered effective only in the $p$--channel. 
The transition from the BCS to the BEC regime can be obtained by tuning the 
strength of the interatomic interaction via a Feshbach 
resonance. In our approach to assess the occurrence of a superfluid phase in the Fermi system 
and to determine its properties, only fermionic degrees of freedom come into play. The lowest 
energy state is a $p_x+ip_y$ superfluid phase in both the BCS and BEC sides. This is in agreement 
with ref. \cite{An61} for the BCS regime and with ref. \cite{Gu05} for the BEC
regime.  In the latter paper  
calculations were performed within  a coupled fermion--boson model for a $p$--wave Feshbach resonance. \par
We expect that the collective modes are strongly damped in the BCS side. Thus, we have studied the low--energy excitation spectrum only in the BEC phase. In this phase the dispersion relations of the collective modes show peculiar features, which might  also be relevant for the transport properties of the system. 
Besides an excitation mode with the usual phonon--like dispersion relation, a gapped mode and a single--particle--like mode can occur.  However, these features are consistent with the Goldstone theorem.  Moreover, when the superfluid Fermi gas is well inside the BEC phase  the chemical potential and the energy per particle approach each--other. This suggests that the pairs of fermions behave as a 
Bose--Einstein condensate with a weak repulsive interaction acting between composite bosons. 
Finally, a comparison has been performed between the dispersion relations of the Fermi--gas collective modes and the corresponding ones of a Bose spinor condensate calculated by means of the 
time--dependent  Gross--Pitaevskii equations. They coincide to a high accuracy. Then, there is a tight 
similarity between a $p$--wave superfluid Fermi gas in the BEC side and a Bose spinor condensate 
for both the equilibrium properties and the dynamical behavior  at low energy. \par 
A final comment is in order. In a mean--field approximation retardation effects, arising from the renormalization of the effective interaction between fermions in a medium, are neglected. In Ref. \cite{Bu09} it has been shown that the energy and momentum dependence of the interaction plays a significant role in determining the $p$--wave pairing gaps in a superfluid Fermi gas. Then, we should expect that, from a quantitative point of view, the inclusion of such effects might affect the results, presented in the present paper, appreciably.  
                                  

\end{document}